\documentstyle[aps]{revtex}

\begin{document}

\title{Possible Self-Organised Criticality and Dynamical Clustering of
Traffic flow in Open Systems}
\author{M. E. L\'{a}rraga, J. A. del R\'{\i}o}
\address{Centro de Investigaci\'{o}n en Energ\'{\i }a,\\
Universidad Nacional Aut\'{o}noma de M\'{e}xico,\\
A.P.34, 62580 Temixco, Mor. M\'{e}xico\\
email: antonio@servidor.unam.mx}
\author{Anita Mehta\thanks{%
Present and permanent address: S N Bose National Centre for Basic Sciences,
Block JD,
Sector
III, Salt Lake, Calcutta 700 091, INDIA, email: anita@boson.bose.res.in}}
\address{Oxford Physics, Clarendon Laboratory,\\
Parks Road, Oxford OX1 3PU, U.K.\\
email: a.mehta@physics.ox.ac.uk}
\maketitle

\begin{abstract}
We focus in this work on the study of traffic in {\it open} systems using
a
modified version of an existing cellular automaton model. We
demonstrate that the open system is rather different from the closed
system
in its 'choice' of a {\em unique steady-state density and velocity
distribution, independently of the initial conditions}, reminiscent of
self-organised criticality. 
Quantities of interest such as average densities and velocities of cars,
exhibit phase transitions between free flow and the jammed
state,
as a function of the braking probability $R$
in a way that is very different from closed systems.
Velocity
correlation functions show that the concept of a {\em dynamical cluster},
introduced earlier in the context of granular flow
is also relevant for traffic flow models.
\end{abstract}

\section{Introduction}

The flow of traffic in congested urban conditions is a subject of
burgeoning
interest in many disciplines at the present time; on the one hand traffic
scientists\cite{nagel},\cite{book} are concerned with the formulation of
models which could study and with luck, ease, congestion problems in the
real
world, while physicists see the subject as an interesting paradigm for
complex systems\cite{jml}. Typically such studies have considered the
behaviour of {\it closed} systems, that is, systems with periodic boundary
conditions which are isolated in the sense that the number of cars is
conserved. Here by contrast we focus on the study of open systems where
nonequilibrium conditions greatly modify the underlying physics, via the
introduction and disappearance of cars at the two ends. Even in the steady
state, we find that the construction of the phase diagram is totally
different, involving as it does an expansion of the phase space, from the
socalled 'fundamental' diagram obtained for the closed version
\cite{nagel}.

The present study is based in the context of the extensively studied model
of Nagel and Schreckenberg \cite{nagel}, \cite{ss}; this involves
four cell-updating steps involving braking, stochastic driver reaction,
and car movement/acceleration. Most studies including stochasticity, as in
the above models,
have been for closed systems with periodic boundary conditions, with open systems
studied mainly
\cite{kk} in deterministic models.

Here we study the effect of open boundary conditions (as occurs in actual
traffic flow) on
a modified Nagel-Schreckenberg model. The modification involves stochastic
changes to the car occurring {\it before} the braking step, to model
the behaviour of an 'anticipatory' driver. Our results include i) a
qualitative
change of the phase diagram, with a {\em unique} steady state for a given
braking parameter $R$, reached from a variety of initial conditions. This is
reminiscent
of ideas of self-organised criticality  (SOC) \cite{BTW}, introduced earlier
in the context of sandpiles.
ii) the manifestation of a peak in velocity correlation functions, at
specific values of $R$,
reminiscent of the dynamical clustering that has been observed in granular
media \cite{bm}.

The NS model has also been developed recently \cite{barl} to show metastable
states
\cite{kr} of very high flow. However we have focused on (a modified
verions of) the simpler, classic NS model to show, in a well-studied context,
that open boundary conditions induce qualitatively new SOC-like behaviour,
as well
as interesting aspects of dynamical clustering. These could, in principle,
be of conceptually useful relevance to more complex models of traffic flow.

This paper is organised as follows. In the first section we present our
model. In the next section we present the results concerning the
steady-state regime in the open systems under study. Lastly we discuss our
results and compare our predictions with observations on real traffic.

\section{The model}

In the real world, traffic flow always occurs in open systems, i.e. those
where cars are always interchanged between some local environment and its
surroundings; thus for example, the number of cars is {\it not }conserved
in
general in any section of a highway. However most studies involving
cellular
automata modelling of such systems have sought to focus on the evolution
of
traffic in closed systems subjected to periodic boundary conditions. In
this
study we seek to model more closely some situations in traffic
flow by looking at systems with {\it open} boundaries where, as in
reality,
the number of cars is not conserved.

We base our model on the Nagel-Schreckenberg \cite{ss} cellular automaton
model, but with the addition of an important modification involving the
order of the operators. Before discussing this, we define the model in its
conventional form:

The model consists of a one-dimensional array of cells each of which can
be
occupied by a car with velocity $v$ between $v_{\min }$ and $v_{\max },$
with $v_{\min }=1$ and $v_{\max }\in \{1,..,5\}$. Subject to the
non-overlapping of cars, the rules for traffic flow are formulated as
follows (we assume that the updating time $t=1$ ):

{\em P}. {\bf Proximity step}

For cars $i$, $i-1$, if $v_{i}+x_{i}\leq v_{i-1}+x_{i-1}$, then
$v_{i-1}^{\prime
}\rightarrow v_{i}+x_{i}-x_{i-1}-1$; else $v_{i-1}^{\prime }\rightarrow
v_{i-1}$, where the primes represent the updated velocities. In words,
this
implies that the driver of a car brakes if the car in front is close
enough
to cause a collision, but not otherwise. Put another way, the driver would
like to be at the maximal possible velocity consistent with the avoidance
of
collisions.

{\em N}. {\bf Noise step}

This reflects the stochastic element which, in the original model, allows
for
the random deceleration of a fraction $R$ of the cars by one unit of
velocity. Thus for example in the case of the $i^{th}$ car, the velocity
$%
v_{i}$ may either stay the same or, if it is part of the randomly selected
fraction $R$ of cars, decrease its velocity
by one unit; thus, $v_{i}^{\prime }\rightarrow v_{i}-1$ (except if $%
v_{i}=v_{\min }$).

{\em M.} {\bf Movement step}

This updates in parallel the positions of the cars; once again for the $%
i^{th}$ car, say, this implies $x_{i}^{\prime }\rightarrow v_{i}+x_{i}$.

{\em A.} {\bf Acceleration step}

This updates in parallel the velocities of the cars by one unit: thus, $%
v_{i}^{\prime }\rightarrow v_{i}+1$ (except if $v_{i}=v_{\max }$).

We emphasise that the above represents the original form of the model in 
\cite{nagel},\cite{ss}, and now proceed to discuss our modification to it,
which involves the order of the operators. Our initial investigations
indicated that the order of rules {\em PNMA }led to several unphysical
configurations, whereas the order {\em NPMA }did not. The reason for this
is
that with the noise being applied {\em after} the proximity step, cars are
unable to adjust to the noise-reduced velocities of the traffic in front.
This could lead to an {\em artificial} jam, arising from the order of the
rules rather than from the real dynamics of the system. Also, importantly, 
 our choice of rules could be said to model the behaviour of 
{\it anticipatory} drivers rather than, as in the case of the {\em PNMA}
ordering,
{\it reactive} drivers.

\section{The steady state in an open system: results and analysis}

In this section we describe both qualitative and quantitative features of
our results for the steady state of traffic flow in an open system, as
described by the model in the preceding section. First of all, we chose
the
system size $L$, randomly generated an initial distribution of car positions
and velocities, and then introduced a car with velocity $v=5$ at the
origin
at every time step. Next we updated the individual car velocities and
positions in accord with the rules of the above model and waited for the
system to asymptote to its steady-state density (where we used the $\chi $
-
squared rule to ensure that this limit was obtained). Finally we recorded
the densities and velocities of cars at different positions for use in our
later analysis. We mention below some of the specific features of our
procedure to ensure convergence to the steady state:

\begin{itemize}
\item  We chose system sizes $L$ from $200$ to $10,000$ units, and found
that although the time required to reach the steady state was enormous as
the system size was increased, the steady-state densities or velocities so
obtained did not vary appreciably. In fact we found that for the really
large system sizes of say $10,000$ units, most of the cars after a
distance
of $\sim 400$ units showed the behaviour trivially to be expected of that
value of $R$, i.e. they were either jammed or free-flowing, and thus no
longer impacted by the initial car. We thus present in this paper only
data
obtained for $L=200,400.$

\item  We varied the rules governing the introduction of the initial car,
for example, choosing to introduce such a car at alternate rather than
consecutive timesteps, and found that this made no significant difference
to
our results.

\item  Lastly we varied the 'seed' configurations to do with initial
densities and velocities on the line, and found that this made absolutely
no
difference to our results. The results presented in this work involve
averages over $1000$ realisations of the experiment.
\end{itemize}

\subsection{Qualitative results}

Our first step is to compare the spacetime diagrams for the case of closed
boundary conditions and open ones, on the former of which one of us has
carried out extensive investigations \cite{mariaelena}. We present below
the
spacetime diagrams for an open system with $R=0.7$ in Fig. 1 and a system
with periodic boundary conditions with the same $R$ in Fig. 2. We note
that
while the specification of an initial density by definition determines the
final density in the closed system (since cars cannot be 'lost' in the
presence of closed boundaries) it does no such thing in the case of the
open
system, where, in the example shown in Fig. 1, the system evolves from an
initially low-density configuration to a jammed state. In some sense we
see
already the signs that the open system 'chooses' its own final density,
while the closed system simply maintains its initially chosen one.

Next we examine the profile of the velocity distribution in the open (Fig.
3a) and closed (Fig. 3b) systems for the {\it same} initial density and
value
of $R$ in both cases. For the closed system, we find a relatively larger
proportion of high-velocity cars persisting even after a long time has
elapsed, compared to the open system, where the number of cars with
velocities greater than 1 decays to zero after an initial transient. (It
is
important to emphasise that the value of the 'most probable' velocity in
each case will depend on $R$). Additionally, while there is a kind of
periodicity that is evident in the case of the closed system, with 'waves'
of cars of a given velocity appearing and disappearing, separated by local
'spurts' in their value, no such phenomenon is observed in the open
system,
where the number of cars with velocity 1 gradually increases with time to
span the system (although there is an interesting rise in the number of
cars
with velocity 2, till its decay to zero at $t \approx 500$). We emphasise
once again that these examples are chosen only to bring out the
differences
between the closed and open systems, and that for example a different
value
of $R$ would result in qualitatively similar but quantitatively different
conclusions.

Next, in Figs. 4 and 5, we show that for the open system, 
initial
conditions involving different densities and different randomly generated
configurations, all converge to the {\it unique} densities and velocities
characterising
the steady state for $R=0.3$ and $0.7$ respectively.
We note that the time required by the open system to converge
to
the steady state is about $10\times L$, where $L$ is the system size \cite
{nagatani}, with the exception of the region around the jamming
transition,
where the transient time can be about $100\times L$. We show, for
comparison, the situation for the closed system in Figure 6; here the
initial densities are maintained, and the value of the steady-state
velocity
depends strongly on the value of the density, {\it unlike the case of the
open system}. Also, in comparison with the open system, the convergence
times are virtually instantaneous.

We see thus that in the open system, {\em arbitrary} initial densities and
velocity distributions {\em evolve towards a unique steady state for a
given 
$R$} characterised by a final mean density and velocity distribution. The
consequences of this apparently simple statement are profound; for example
the fundamental flux vs. density diagram obtained in the case of the
closed
system \cite{eis} for a given value of $R$ {\em collapses to a point} in
the
open system, since there is only one possible value of density $\rho$ and
velocity $v$ in the latter case.

We discuss this unique 'selection' by the open system of steady-state
densities and velocities later, but for the present, simply assert that
this
convergence enables us to work with average densities and velocities
(obtained by averaging over time, in the steady state, as well as space,
and
finally over different initial configurations and noise realisations of
the
system) in the next subsection.

\subsection{Densities, velocities and correlation functions: a
quantitative
analysis}

We next present and interpret quantitative results on average velocities
and
densities of cars in the steady state, in addition to examining their
fluctuations via correlation functions. In Fig. 7a, the mean density and
velocity for systems of size $L=200$, $400$ are plotted as a function of
$R$%
. As is evident, the curves are coincident, reflecting our contention that
the steady state obtained in our work is not system-size dependent beyond
about $L=200$. We see strong evidence of a phase transition which arises
around $R_{c}\sim 0.55$, $\rho _{c}\sim 0.55$. (These numbers are obtained
from an analysis of $\frac{d\rho }{dR}$ vs $R$ , which is shown in Fig. 7b 
;
we will have more to say about the latter graph and its implications later
on).

We notice that the density curve is a smooth S-shaped function while that
for the velocity is a smooth inverse S-shaped function. Their intersection
indicates the likely neighbourhood of the phase transition observed
between
regions of low $\rho $ and high $v$ ('freely flowing traffic') on the one
hand, and regions of high $\rho $ and low $v$ ('congested' or 'jammed'
traffic) on the other. Earlier work on closed systems seems to categorise
phase transitions in traffic flow as being of first order \cite{eis} but
we
are unable to state this definitively in the context of our finite-size
investigations on open systems. In particular the 'selection' by the
system
of steady state densities and velocities for a given value of $R$ is
rather
reminiscent of the phenomenon of self-organised criticality, \cite{BTW},
where the system organises itself into a unique state for a given value of
a
parameter. On this basis $R$ would seem to be analogous to a
temperature-like variable which then determines the density $\rho$, whose
thermodynamic analogue is the system energy.

However, a deeper examination of this issue is relegated to future work,
as
for example the shape of $\frac{d\rho }{dR}$ vs $R$ (analogous to the
temperature dependence of the specific heat of a thermodynamic system)
depicted in Fig. 7b, could equally well represent a second-order
transition
for a finite system, or for example a kind of lambda transition,
reminiscent
of the first-order transition in glassy systems \cite{sidnagel}.

We now turn to the discussion of fluctuations via the analysis of
correlation functions. Clearly the $<\rho _{x}\rho _{x^{\prime }}>$
correlation function is not very informative at least in its 'bare' version
(i.e. where its value is either $0$ or $1$ at a site); on the other hand,
the $<v_{x}v_{x^{\prime }}>$ correlation function is meaningful.
 (Since we look only at the steady
state behaviour here, time correlation functions such as
$<v_{t}v_{t^{\prime
}}>$ are likewise not meaningful). In Fig. 8 we present the
behaviour of this as a function of position, for different values of $R$.
 We note that the behaviour is generic,
with well- defined first and second neighbour 'shells', particularly for
values of $R$ well away from the transition point. Additionally, we remark
on the specific meaning of such {\em dynamical} correlations; in analogy
with earlier work on granular flow \cite{bm}, we define a {\em dynamical
cluster} for a given $R$ as being the number of sites which
are
within the first shell of the velocity correlation function . {\it The
physical
import of a dynamical cluster is that it reflects the range over which
cars
are correlated in their velocities; we observe that the size of a
dynamical
cluster increases as $R$ decreases}. In other words, as fewer cars face
random obstacles, more and more of them develop velocity correlations,
i.e.
they begin to 'move together' in clumps. Returning to the analogy with
granular flow, this mirrors the situation found in earlier work \cite{bm}
where a decrease in external perturbations applied to a granular system
causes an {\it increase} in the size of a typical dynamical cluster of
grains.

\section{Discussion}

We have examined traffic flow in open systems, and found that the nature
of
the phase diagram is completely altered with respect to the more usual
case
of periodic boundary conditions. In particular, the fundamental diagram of
flux versus density as a function of the parameter $R$ presented recently
for closed systems by Eisenblatter et al \cite{eis} collapses to a point
in
the case of an open system; thus, at a given $R$, traffic flow in an open
system is characterised by a {\em unique density and velocity
distribution,
independently of initial conditions}.

This unusual and very robust feature leads us to suggest some
thermodynamic analogies for the key quantities in traffic flow in open
systems: thus, for example {\it the thermodynamic analogues of density $\rho$
and braking
probability $R$ are respectively energy and temperature}.
Following this line of reasoning, we speculate that traffic flow in open
systems could either be a paradigm of self-organised criticality, or on
the
other hand be representative of a first-order phase transition in a finite
system. The transition in question, that between jammed and free flow,
appears to be characterised by a discontinuity in the analogue of the
specific heat as a function of $R$, i.e. $\frac{d\rho }{dR}$ plotted vs
$R$
shows a lambda-transition which could be characteristic either of glassy
behaviour of indeed of self-organised criticality. 

Various special cases of traffic flow modelled by cellular automata have
been examined and found to exhibit self-organised criticality
\cite{nagatani}%
; for example, the case of the outflow region of a big traffic jam under
cruise control conditions \cite{NP} was found to exhibit this. However, we
reiterate that our work is to our knowledge the first to investigate the
specific issue of the phase diagram as a function of the braking
probability 
$R$ {\em under the most general conditions}. Our striking findings
regarding
{\it the selection by the system of a unique density and velocity
distribution}
for arbitrary initial conditions suggest that it may well be a rather
general paradigm of self-organised criticality, though future work is in
progress to investigate this.

Lastly, we mention that in recent experimental work \cite{kr} there has
been
a suggestion that in addition to the transition between jammed and free
flow, there could be a transition to 'synchronised' flow where cars
neither
move freely, nor stay jammed, but continue moving by synchronising their
velocities. Our findings with regard to the {\em dynamical cluster}
mentioned in the earlier section appear to be in accord with this, in that
dynamical clusters, as discussed earlier in the context of granular flow 
\cite{bm}, are clusters whose constituents are strongly correlated in
their
velocities.
We hope to explore some of these issues  elsewhere.
\section{Acknowledgements}

AM acknowledges the generous hospitality, over many visits, to the Centro
de
Investigaci\'{o}n en Energ\'{i}a in Temixco, where a large portion of this
work was carried out. This work was partially supported by DGAPA-UNAM
under
project IN117798.
We are very grateful to Subodh Shenoy for a careful reading of
the manuscript.

\section{Figure Captions}

 Figure 1. Spacetime diagram for traffic flow in an open system
 corresponding
 to a braking probability $R=0.7$, and starting with an initial density
 $\rho
 _{i}=0.2$.
 
 Figure 2. Spacetime diagram for traffic flow in a closed system
 corresponding to a braking probability $R=0.7$, and a density $\rho =0.2$.
 
 Figure 3. Profile of the velocity distribution for traffic flow in a) an
 open system and b) a closed system corresponding to a braking probability
 $%
 R=0.7$, and starting with an initial density $\rho _{i}=0.2$.
 
 Figure 4. Plots of the time evolution of the a) density and b) average
 velocity of traffic in an open system for two initial densities $\rho
 _{i}=0.2$ and $\rho _{i}=0.7$, and braking probability $R=0.7$. Both
 initial
 conditions evolve to a single density characteristic of the jammed state.
 
 Figure 5. Same as Figures 4 but with braking probability $R=0.3$; the
 final
 state is, as expected, characteristic of free flow in this case.
 
 Figure 6. Evolution of the time dependent averaged velocity for closed
 systems with two initial densities $\rho =0.2$ and $\rho =0.7$, and
braking
 probability $R=0.3$. In this case we notice that the final state depends
 strongly on the (initial) values of the density.
 
 Figure 7. a) The 'fundamental diagram' of traffic flow in open systems;
the
 free-flow to jamming transition occurs in the vicinity of the intersection
 of the density and velocity curves as a function of braking probability
 $R$.
 b) plot of $\frac{d\rho }{dR}$ vs $R$; note the strong resemblance to the
 lambda transition in glassy systems. Triangles indicate the results for a
 system of length $L=200$ while open circles indicate the data for a system
 size $L=400$.
 
Figure 8. Velocity-velocity correlation functions $<v_{x}v_{x^{\prime }}>$
 corresponding to a range of different values of the braking probability
 $R$.

\end{document}